# Report on the Workshop on Accelerator R&D for Ultimate Storage Rings

Huairou District, Beijing, China

Oct 30-Nov 1, 2012

R. Hettel, SLAC National Accelerator Laboratory, Menlo Park, CA, USA

Q. Qin, Institute of High Energy Physics, Beijing, China

Report revision date: March 7, 2013

# Contents

# Executive Summary

The Workshop on Accelerator R&D for Ultimate Storage Rings was held on October 30 to November 1, 2012, at the Hongluoyuan Hotel in Huairou District, north of Beijing, near the site of a new campus for the Chinese Academy of Sciences and potentially a future state-of-the-art storage ring. IHEP in Beijing hosted this international workshop because of it is seeking special support from the national funding agency to conduct R&D related to the new 5-GeV, 1.2 to 1.5-km circumference storage ring-based light source. About 60 accelerator physicists and engineers from several international light sources attended. The Chairmen, Local Organizing and International Advisory and Science Program Committees are shown in Appendix A.1. The workshop website is found at http://usr2012.ihep.ac.cn/.

It is well known that the interest in realizing the next generation of diffraction-limited, so-called "ultimate" storage ring (USR) light sources is growing and several laboratories, including SLAC, SPring-8, ESRF, now IHEP/Beijing and possibly other laboratories are considering implementing them in their strategic plans for the next decade. The design of these machines, which have electron emittances of < ~100 pm in both transverse planes, have been discussed in the last two ICFA Future Light Source Workshops and have been the topic of study by individual groups. It is acknowledged that R&D is required in various accelerator physics and engineering areas before such machines can actually be implemented, especially as emittance is reduced towards 10 pm. These rings will most likely use compact magnet and vacuum chamber technology similar to that being developed for the MAX IV storage ring, now under construction.

The purpose of this workshop was to bring together accelerator experts from diverse light source facilities having common interest in developing these new sources to focus on accelerator physics and engineering challenges for USRs and to identify areas requiring R&D. The workshop was organized with an opening introductory session that included presentations on the science case and performance goals for diffraction-limited storage ring light sources. This was followed by several topical sessions, interspersed with discussion sessions, organized to identify issues in lattice design, accelerator physics, injection, accelerator engineering, instrumentation and feedback systems, and insertion devices relative to the present state-of-the-art. While possible avenues of solution were discussed in some cases, the identification of these issues was the primary purpose of the workshop.

Since the science case is still being developed, and no hard requirements for USR performance are yet being requested by the scientific community, there are many more questions being asked than answers being provided. In the meantime there are two branches of development of the next generation of storage ring light sources: those involving replacement of existing lattices, imposing constraints that limit the reduction of emittance to the order of 50-100 pm-rad (e.g. ESRF and SPring-8), and larger greenfield machines that might push emittance to 10 pm-rad or less having beam parameters that may enable FEL operation.

The following report includes general discussion of USR performance goals and summary notes on the issues and R&D topics identified by the topical session working groups. This information may help in developing a comprehensive R&D plan for USRs in the future, and obtaining support from funding agencies for carrying out the plan. It is noted that further development of the science case and the subsequent definition of performance requirements for USRs will be crucial for this effort.

# 1. Introduction and Workshop Charge

Storage ring-based light sources will continue to play a vital role in X-ray science into the future since they offer beam properties that are complementary to FEL sources. Ring-based sources provide highly stable photon beams having low peak brightness with high average brightness and high pulse repetition rate, photons that do not over-excite or damage samples the way those from FELs do, and they serve a large number of diverse users simultaneously. There are now emerging scientific applications and experimental methods that would greatly benefit from ring-based sources having much higher brightness and transverse coherence than present or near-future storage ring facilities – storage rings having electron emittance of ~100 pm-rad or less in both transverse planes – on the scale of the diffraction limited emittance for hard X-rays. Several institutions world-wide are now including the prospect of building diffraction-limited "ultimate" storage rings (USRs) in their 10-year development plans. These machines push the state-of-the-art for storage ring accelerator and photon beam line design, presenting many significant challenges that must be addressed with R&D.

**Charge for the workshop:**

- Survey conceptual designs and compare the performance goals for USRs worldwide
- Identify technical challenges and R&D requirements associated with:
    - Ring dimensions and lattice design
    - Collective effects, impedances and lifetime
    - Injection methods
    - Accelerator component and system design (magnets, vacuum chambers, instrumentation, feedback systems, etc.)
    - Beam stability
    - Insertion devices and damping wigglers
- Prioritize R&D topics and define critical studies that should begin imminently

We suggested that each talk include (but not necessarily limited to) the following:

- A concise description of the topic being presented, including design goals, present state –of-the-art performance (if applicable), and a statement of the associated challenges for reaching USR implementation goals.
- A concise description of the methods and principles and any demonstrated results associated with the technology being presented and how they could help reach USR implementation goals.
- If applicable, a statement of any R&D (and an estimate of associated time and manpower if possible) needed to realize the technology being presented.

**Note on other USR workshops:**

We note that there have been workshops in the past addressing the design of diffraction-limited ring-based light sources - storage rings and ERLs – and their science applications. Among these are the DOE/BES Future Light Sources Workshop in 2009 [1], the ICFA Future Light Source Workshops at SLAC in 2010 and Jefferson Lab in 2012 [2, 3], and a series of science workshops held at Cornell in June 2011 [4]. At the time of the Huairou Workshop, another science workshop was scheduled at SPring-8 in December 2012, and has since taken place. In addition, an informal USR study group was formed in the US with participants from ANL/APS, BNL, LBNL/ALS and SLAC/SSRL which addressed beam line and X-ray optics issues as well as accelerator issues. The accelerator topics discussed in these meetings were basically mirrored in the Huairou Workshop but with more focus on USR-specific issues. The resulting R&D topics identified at Huairou are, not surprisingly, similar to those from earlier workshops but they are more specific, having being informed from more detailed USR design studies that have taken place over the last few years, and especially over the last year as the possibility for actually building the next generation of storage ring light source has become more real. We also note that very similar accelerator topics are being considered by the very low emittance linear collider damping ring design community (e.g. the CERN-sponsored Low E Ring workshop series [5]) and the possibility of a future merging of efforts for these two applications is foreseen.

## 2. USR Science and Design Goals

### 2.1 Science Case

The science case for diffraction-limited light sources, including USRs and ERLs, is being developed within the international light source community [4]. In the case of USRs, the science case has yet to be fully articulated in a way that clearly defines accelerator design goals beyond just "increasing brightness and coherence as much as possible" with reasonable cost and practical accelerator designs. At the moment, science applications are presently aimed at using increased brightness and coherence for nano- and meso-scale science using techniques that include:

- diffraction of single nano-objects
- coherent diffraction imaging (CDI), including lensless imaging (e.g. ptychography) of meso-scale structures (3~5 nm)
- X-ray photon correlation spectroscopy (XPCS): dilute samples, better time-resolution

The possibility of more dramatic performance from diffraction-limited rings, such as high repetition rate short bunches and perhaps even FEL operation, is stimulating the community to define related applications. The complete science case for future rings will continue to develop as the performance potential and related implementation requirements are more fully understood by the accelerator community.

### 2.2 Accelerator Performance and Design Goals

**High brightness**

The spectral brightness envelopes for existing and near future storage ring light sources are depicted in Figure 1. It can be seen that existing modern 3$^{rd}$ generation machines have brightness of $10^{21}$ (ph/s/mm$^2$/mrad$^2$/0.1%BW) or less, a value that will be pushed to the mid-$10^{21}$ by the new NSLS-II and MAX IV rings in the near future. The brightness goal for future rings is at least an order of magnitude greater than this value.

Average spectral brightness $B_{avg}(\lambda)$ is characterized by

$$B_{avg}(\lambda) \propto \frac{N_{ph}(\lambda)}{(\varepsilon_x \oplus \varepsilon_r(\lambda))(\varepsilon_y \oplus \varepsilon_r(\lambda))(s \cdot \%BW)}$$

where spectral flux $N_{ph}(\lambda)/s$ is proportional to ring electron current $I_{e^-}$, $\varepsilon_x$ and $\varepsilon_y$ are transverse electron emittances, and $\varepsilon_r$, added in quadrature with the electron emittances, is the diffraction-limited photon emittance at wavelength $\lambda$ given by $\varepsilon_r \cong \lambda/4\pi$. This formula assumes that the orientations of emittance phase space ellipses for electrons and photons are matched (i.e. $\sigma_x/\sigma'_x = \sigma_y/\sigma'_y = \sigma_r/\sigma'_r \cong L_{ID}/2\pi$, where $\sigma$ and $\sigma'$ indicate RMS size and divergence of Gaussian beams, $\sigma\sigma' = \varepsilon$, where, for an undulator of length $L_{ID}$, $\sigma_r \cong (2\lambda L_{ID})^{1/2}/4\pi$, and $\sigma_r \cong (\lambda/2L_{ID})^{1/2}$). Higher brightness can be reached by increasing electron current or by reducing transverse emittance. Since practical stored beam currents are limited to present levels of a few hundred milliamps by photon power issues in the X-ray beam line and experiment, the path to future high brightness rings is to reduce electron emittance. Since most light sources already operate with vertical emittances near the diffraction limit for mid-keV X-rays by minimizing horizontal-vertical emittance coupling, the horizontal emittance must be reduced.

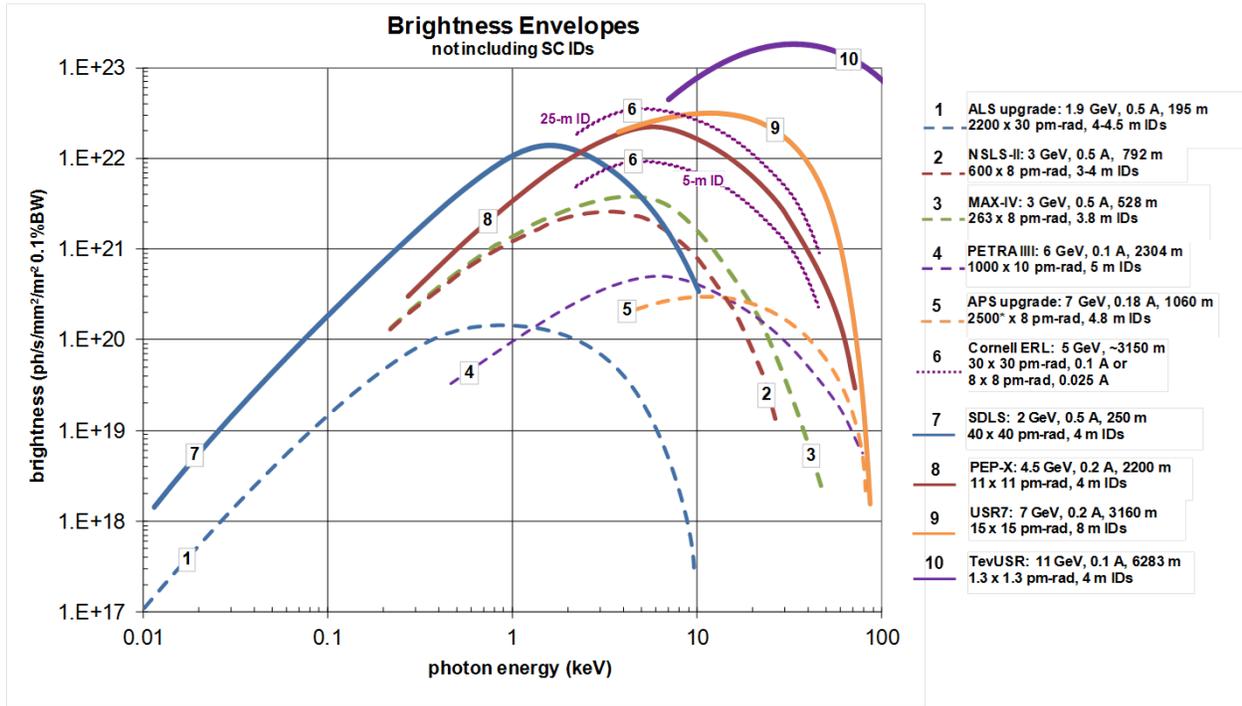

**Figure 1.** Spectral brightness envelopes for existing storage ring light sources and future USRs having three dominant X-ray spectral ranges: ≤ 2 keV, 2-20 keV, ≥20 keV. Brightness curves for the proposed Cornell 5-GeV ERL are included for comparison.

Natural horizontal electron emittance $\varepsilon_{x0}$ in a storage ring is characterized by

$$\varepsilon_{x0} = \frac{C_q E_{e-}^2}{J_x} \theta_B^3 F_{latt}$$

where $E_{e-}$ is the electron energy, $\theta_B$ is the bending angle of the dipole magnet making up a unit cell in the lattice, $J_x$ is the horizontal damping partition, $F_{latt}$ is a value dependent on lattice type, and $C_q$ is a constant. For a given cell type with fixed dipole length, $\theta_B$ is reduced by increasing the number of dipoles in the lattice, thereby increasing ring circumference C and yielding an approximate emittance scaling given by

$$\varepsilon_{x0} \propto \frac{E_{e-}^2}{J_x C^3} F_{latt}$$

Assuming that ring energy $E_{e-}$ is approximately fixed by the desired X-ray spectrum (although it is a variable within limits), the primary path to low emittance is to increase the number of dipoles in a lattice having small $F_{latt}$ and to maximize $J_x$ (although in practice $J_x$ can only be modified by a factor of 2 or less). Other factors influencing emittance include emittance growth due to intrabeam scattering (IBS) of electrons within small bunches, emittance growth due to self-generated coherent synchrotron radiation (CSR) from very short bunches, and the use of damping wigglers to reduce emittance.

While most 3rd generation storage rings use double-bend achromats (DBAs) or triple-bend achromats (TBAs) for their lattices, It was recognized years ago that a higher number of bends could be incorporated into "multibend achromats" (MBAs) as a way to reduce emittance [6]. MAX IV will be the first ring to incorporate seven-bend achromats (7BA, Figure 2), reaching 250-pm-rad emittance at 3 GeV with a 528-m circumference. The USR designs used for the brightness plots in Figure 1 all use similar 7BAs. Examples include the 2.2-km, 4.5-GeV PEP-X ring having an emittance of 11 pm-rad, and the 9-GeV TevUSR that could be built in the 6.28-km Tevatron tunnel at Fermilab having an emittance of 1-3 pm-rad. The number of bends in the achromat is limited by available space; for this reason, the lattice upgrade for SPring-8, having a cell-length constraint imposed by their existing DBA ring geometry, is planned to be 6BA; on the other hand, the replacement of the ESRF DBA lattice is planned to be with a "hybrid 7BA" lattice having the same cell length (Figure 3).

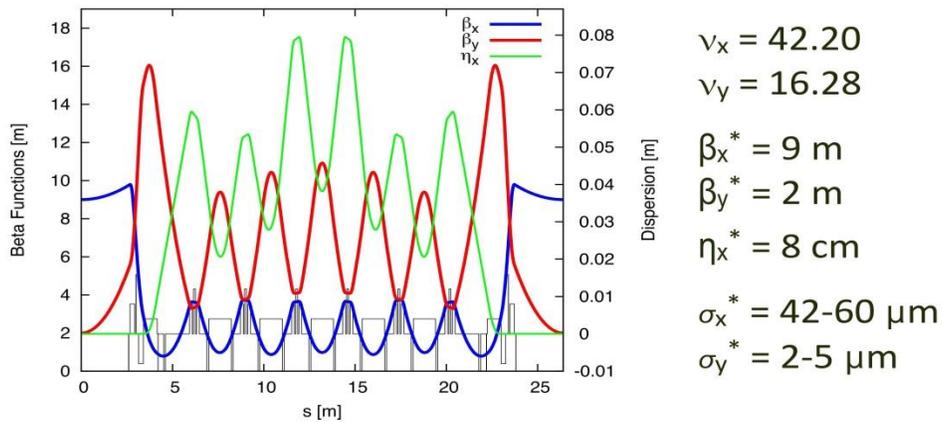

**Figure 2.** MAX IV 7-bend achromat (7BA).

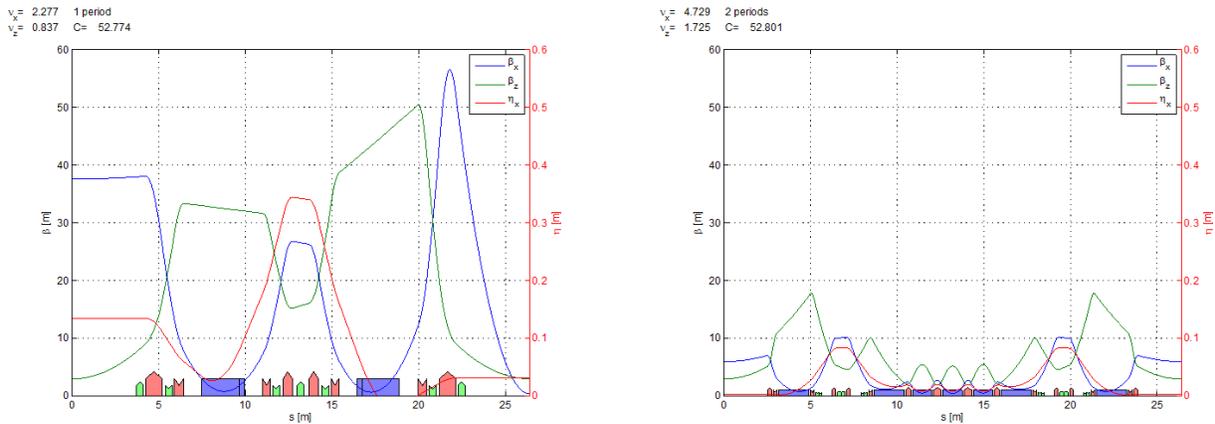

**Figure 3.** The ESRF plans to replace its DBA lattice (left) with a "hybrid 7BA" lattice (right) that provides high dispersion points for sextupoles in order to reduce their gradients.

Design challenges associated with such low-emittance designs include achieving sufficient dynamic aperture in the lattice design and the engineering of very high quality and compact magnets and vacuum chambers. In some cases, on-axis swap-out injection [7] may be required to accommodate small dynamic aperture. These are discussed more completely in Section 3.

**High coherent fraction**

Closely related to beam brightness is the fraction of photons that are transversely coherent. A high coherent fraction serves to maximize the achievable performance for the experimental methods given in Section 2.1.

Coherent fraction $f_{coh}$ is characterized by

$$f_{coh}(\lambda) = \frac{\lambda/4\pi}{(\varepsilon_x \oplus \lambda/4\pi)} \cdot \frac{\lambda/4\pi}{(\varepsilon_y \oplus \lambda/4\pi)}$$

where $\varepsilon_r = \lambda/4\pi$ is the diffraction-limited emittance for wavelength $\lambda$, and again assuming matched emittance phase space orientations. Figure 4 shows the diffraction-limited emittance as a function of photon energy and what emittance regions are accessed with present and future storage ring light sources.

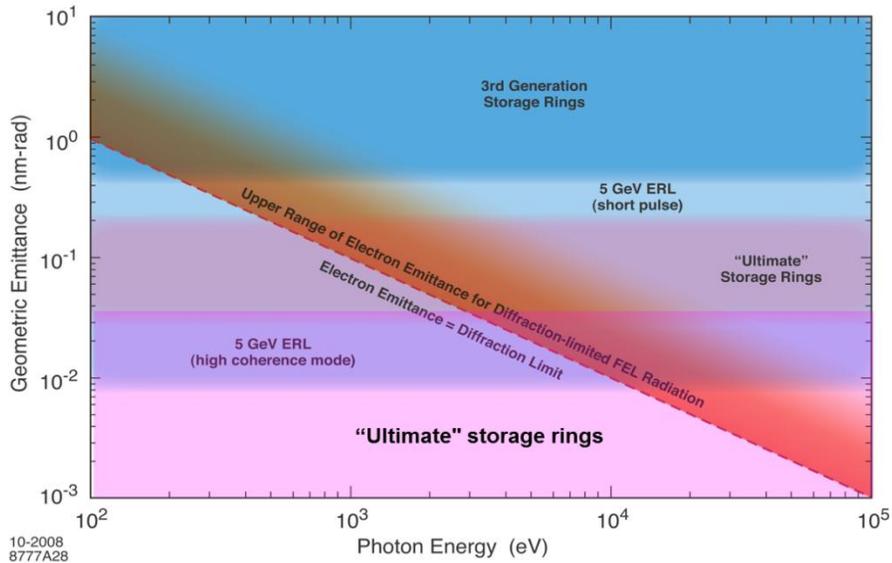

**Figure 4.** Diffraction-limited emittance as a function of photon energy. The diffraction-limited emittance for 12-keV photons (1 Å) is 8 pm-rad.

It can be seen that when $\varepsilon_x$ and $\varepsilon_y$ are at the diffraction limit $\varepsilon_r$, $f_{coh}$ is 25%. Electron emittance significantly smaller than the photon emittance is needed to approach a coherent fraction of 1 (Figure 5).

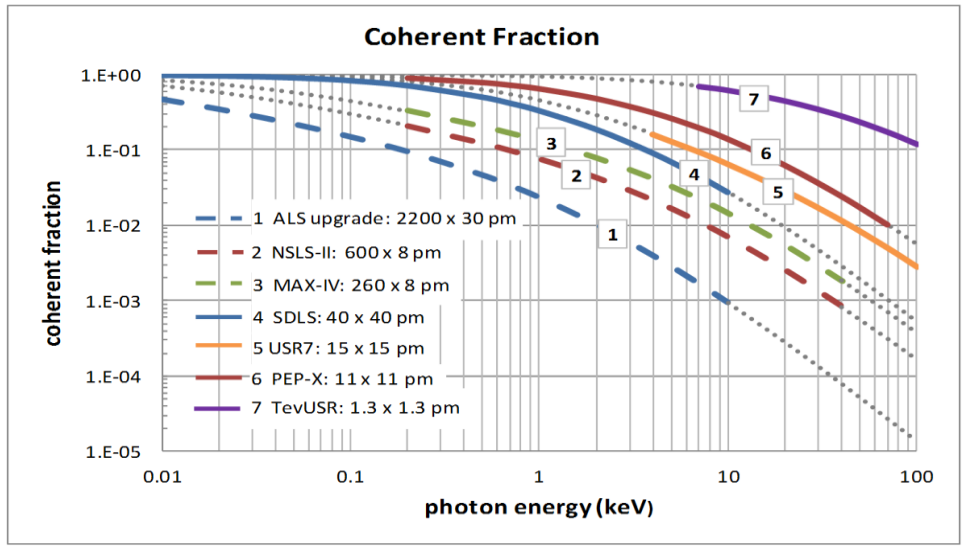

**Figure 5.** Coherent fraction for future rings.

**Optimized ring configuration**

For a green-field USR, the optimization of storage ring parameters is a complex process that could yield a range of solutions depending on factors that include the spectrum of interest, the necessary beam emittance, the available space, the number of insertion device source points and their straight section lengths, and any advanced performance requirements (e.g. short bunches, etc.), and almost certainly the most significant factor: available funding.  As mentioned earlier, this optimization will depend on the science requirements and, since it is unlikely that there is any sharply defined threshold in performance beyond which new science would be enabled, the optimization is likely to be "soft", driven primarily by cost.  For example, the science community must decide whether a 10-pm-rad machine having 2-km circumference is worth the substantially higher investment than needed for a 100-pm, 1-km ring.  On the other hand, a performance threshold does likely exist if X-ray FEL operation is to be realized – emittance on the order to 10 pm-rad, the need for peak bunch current higher than normally found in storage ring, and the need for very long straight section(s) for the FEL undulator(s) (on the scale of 100 m).

Included in the optimization for very large rings is the possibility of consolidating beam line source points in certain regions of the ring, leaving other regions without beam lines, as way to minimize experimental hall construction costs and maximize operational support efficiency.

**Possibility for "round" beams**

With very small horizontal emittances approaching the diffraction limit for X-rays, USRs could be operated very effectively with "round" or "quasi-round" beams, as opposed to the very flat beams generally found in 3rd generation rings characterized by much larger horizontal emittance but nearly diffraction-limited vertical emittance.  More accurately described, these "round" beams are nearly symmetric in 4-dimensional (size and divergence) phase space. Transversely symmetric, low emittance beams are often more advantageous than significantly asymmetric beams, even those having comparable total emittance. Advantages include more optimal coupling to some optics (e.g. circular zone plates) and detectors (e.g. detectors with arrays of square pixels) or in coupling to experiment and detector where beam symmetry can simplify experimental equipment and detector boundary conditions, instrument resolution functions, etc. The round electron source also could enable the use of high performance insertion devices having small horizontal and vertical apertures (e.g. the Delta undulator [8] operating in helical mode where the on-axis intensity of unwanted harmonics is suppressed). Perhaps more fundamentally, the symmetry in the transverse coherence of such a beam facilitates measurements that require or exploit coherence lengths or coherent fractions that are comparable in both transverse directions, and that might exploit, for example, optics-less imaging configurations.

Methods to create round beams include operating with equal horizontal and vertical tunes, using skew quadrupoles, solenoids and other methods. These methods need further study to determine which is best.

**High beam current constancy**

$3^{rd}$ generation light sources are already benefiting by the high level of beam current constancy afforded by top-up injection. USRs are likely to have lifetimes on the order of 1 h, so frequent injection will be required to maintain beam current constancy to better than 1%. This requirement will impose design challenges for injectors on rings needing on-axis injection, as discussed in Section 3.

**Possibility for high rep-rate picosecond bunches**

The natural bunch length for the very low emittance USR lattices tends to be fairly short – on the scale of 10 ps RMS. This bunch length can be reduced to a few picoseconds using a harmonic RF cavity together with the ~500-MHz nominal RF system, or to the picosecond level by using a higher frequency, higher voltage RF system (~1.5 GHz) in place of the typical 500-MHz system, or by using a combination of frequencies operating in beat-frequency mode to create alternating long and short bunches (as proposed by SPring-8), or by pulsed RF or other methods. The availability of high repetition-rate picosecond bunches would enable MHz-scale pump-probe measurements of materials dynamics occurring on time scales of a few picoseconds or more, a temporal range not accessible with pulsed linac FELs. These types of measurements are presently being pursued at the APS which is in the process of installing superconducting crab cavities to create the short bunches in a localized region of the ring.

**Possibility for advanced performance capabilities**

While USRs will have unsurpassed brightness and coherent fraction in the storage ring light source community, there are possibilities that other performance capabilities could be realized. These include the possibility of propagation sub-picosecond bunches from a linac injector for several turns around the ring, providing a burst of high repetitions rate short pulses, and the possibility of operating in non-standard lattice configurations to create "tailored bunches" [9], bunches that have different properties tailored to different users. But perhaps the most compelling possibility is that for lasing at soft X-ray energies using single-pass FEL undulators located in switched bypasses [10], or potentially lasing at hard X-ray energies in X-ray FEL oscillators [11]. Both of these implementations would likely use transverse gradient undulators (TGUs) to accommodate the relatively large energy spread of the storage ring beam.

In general, it is a challenge for USR designers to push the limits of performance in an attempt to make them more "FEL-like". Figure 6 illustrates various photon beam properties as a function of pulse duration for FEL and ring-based light sources and the directions that USR performance could go (indicated with pink arrows) with combinations of bunch compression and lasing capabilities.

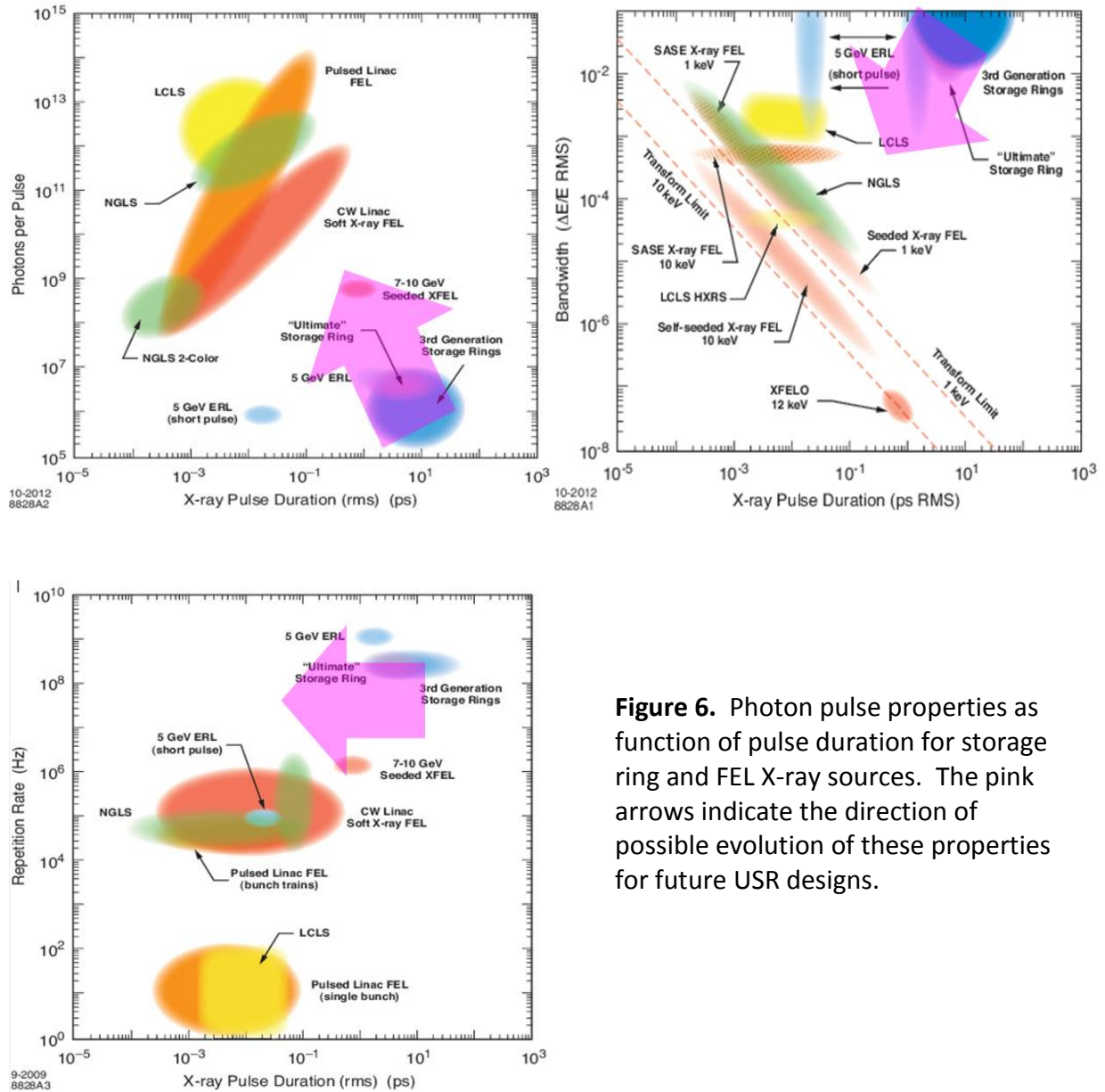

**Figure 6.** Photon pulse properties as function of pulse duration for storage ring and FEL X-ray sources. The pink arrows indicate the direction of possible evolution of these properties for future USR designs.

# 3. Workshop Session Notes

The following sections contain notes on the primary observations made and issues identified during workshop session presentations and related discussions. Summaries of the presentations themselves are not presented. The presentations can be found at the Workshop website: http://indico.ihep.ac.cn/conferenceOtherViews.py?view=standard&confId=2825

or by contacting one of the conference chairmen: Qing Qin, IHEP (qinq@ihep.ac.cn) or Robert Hettel, SLAC (hettel@slac.stanford.edu).

The identified R&D topics for each session topic are presented in Section 4.

## 3.1 Accelerator Lattice Design

Questions for consideration in these sessions included:

1. What approaches are being used to optimize parameters for USRs (e.g. emittance, current, energy, circumference, straight section lengths, Twiss parameters, etc.) and what are the conclusions so far?
2. Is there an optimal M for a greenfield MBA lattice?
3. What are ring geometries and lattice configurations that optimize photon beam line layout? Should hybrid lattices be used to consolidate beam lines in very large rings? Are there novel "non-circular" geometries (e.g. using chicanes, etc) to optimize beam line layout?
4. Should very long straight sections be included to provide space for beam manipulation components, FEL implementations, etc.?
5. How can dynamic aperture and momentum acceptance be maximized?
6. Should emittance reduction be limited by the requirement to inject off-axis?
7. What methods are best for producing near-round beams in IDs?
8. Can USR lattices be operated in isochronous mode to propagate short bunches for a few turns without sacrificing stored beam emittance? Can the lattice be compatible with ERL operation, including the ability to tune individual straight section parameters?
9. Can lattices have an "emittance knob" permitting evolution to the diffract limit over time?
10. Can USRs accommodate IDs having small horizontal aperture (e.g. vertically oriented IDs, small-aperture DELTA undulators, etc.)?
11. Do tracking and simulation codes need development?
12. What studies can be performed on existing storage rings?
13. What R&D is needed before an actual USR is built?
14. What should be the emittance goal for a 1-1.5 km ring in the near future?

Not all questions were addressed in the Workshop, but they remain for future consideration.

Presentations given in this session are listed in Appendix A.2. The following items were noted in the session and discussion periods:

**Lattice design:**

1. Good progress made in low emittance lattice design using multi-bend achromats (MBAs). It is noted that the design profits from a positive feedback cycle: many small cells keep dispersion low and thus allow reduced apertures, which in turn allow higher gradients and thus shorter magnets and more cells per length.

2. For a given circumference C, the space available for MBAs, and thus the number of cells M in an achromat, is limited by the number of straight sections and their lengths.

3. Is local control of beta functions needed? Some think not.

4. Can USRs accommodate IDs having small horizontal aperture (e.g. vertically oriented IDs, small-aperture DELTA undulators, etc., with, for example, on-axis injection)?

**Lattice optimization:**

1. The general issue of optimizing ring parameters (e.g. E, $\varepsilon$, C, $\beta_{x,y}$, RF, straight sections, etc.) based on targeted spectral brightness, coherence, special operating modes (e.g. short bunches, lasing) was not specifically addressed. However it has been shown that minimum $\varepsilon$~10 pm with IBS @ 0.5 nC/bunch is found for E ~ 5 GeV for ~1.5-km USRs. A more conservative emittance goal for such a ring may be prudent in near future designs.

2. Determining the optimal length for straight sections depends on factors such as the user need for multiple IDs in one straight section, providing room for future and possibly unforeseen components, the trade-off of achromat and straight section length, etc. Providing the possibility of 2 IDs in a straight section could be a cost-effective way to consolidate beam lines in sub-sections of large rings.

3. Can a quality factor be defined to gauge lattice optimization in terms of emittance normalized to the energy, circumference and total length for straight sections?

4. Electron-photon phase space matching is a design criterion for maximum brightness.

5. Consider lattices that consolidate beam lines in large rings (cost, operational ease), possibly using non-circular geometries and/or hybrid lattices.

6. Photon scientists would like to understand the range of performance possibilities and trade-offs, perhaps illustrated with a performance envelope in 3 dimensions for a given ring energy: beam emittance, beam current and bunch length.

**Dynamic aperture:**

1. Obtaining adequate dynamic aperture and beam lifetime should be possible by reducing resonance driving terms and using high order multipole magnets. Localized cancellation of resonance driving terms up to $4^{th}$ order over the length of one arc has been achieved in the 11-pm-rad, 7BA lattice design for the 6-arc, 4.5-GeV PEP-X ring.

2. ESRF has optimized dynamic aperture using a "hybrid 7BA" lattice that includes dispersion bumps to reduce sextupole strength, small dispersion in central dipoles, longitudinal dipole gradient (Figure 3), providing a way to optimize the non-linear lattice that is complementary to the normal methods. Note: this presentation was given in the Accelerator Physics session.

3. Multi-objective genetic algorithms (MOGAs) are a powerful tool for optimization.

**Damping and Robinson wigglers:**

1. Damping wigglers can reduce emittance by a factor of 2 or more and are especially effective for counteracting emittance growth due to IBS. On the other hand, they produce large photon power needing special absorber designs, require large RF power, increase electron energy spread and introduce non-linear fields that can reduce dynamic aperture. They are less useful for high energy rings.

2. Robinson wigglers might be used to reduce emittance by increasing the horizontal damping partition for some lattice designs, but this has yet been confirmed. It is noted that these wigglers are placed in non-achromatic straight sections which may constrain USR lattice design and possibly prevent the reduction of emittance by dispersion leakage into achromatic straight sections.

3. The decision to use damping wigglers or not should be made as part of the optimization of ring parameters, including beam energy, current, emittance, RF frequency, dipole field strength, circumference and other parameters.

**Round beams:**

1. Explore and study different solutions, including equal tunes, skew quads, solenoids, vertical dispersion, vertical wigglers, etc. Identify sites where to make tests.

**Coupling correction** (in flat beams mode):

1. Well established know-how and procedures; seems already OK for USRs.

2. It is desirable to control $\varepsilon_y$ without adding (too much) coupling, including "white noise" excitation.

**Momentum compaction:**

1. USR lattices typically have low momentum compaction $\alpha$, leading to shorter bunches, increased impedance-related and stability issues.

2. Can chicanes be used to increase $\alpha$?

3. Should $\alpha$ be readily variable by design to enable short bunch lattice tunings?

Members of the Lattice Design Working Group included A. Franchi, D.H. Ji, L. Liu, A. Nadji, H. Ohkuma (co-chair), Y. Shimosaki and G. Xu (co-chair).

## 3.2 Accelerator Physics

Questions for consideration in these sessions included:

1. What are collective effect and IBS issues for USRs?
2. What are narrowband and broadband impedance limitations for USRs?
3. How can CSR effects be mitigated?
4. Are ions an issue?
5. How is lifetime maximized?
6. How can short bunches be generated and for how many turns can they be circulated in the ring?
7. Can ~200 peak amps be achieved for lasing? Can beam manipulation be used?
8. What are beam manipulation methods and applications (emittance exchange (RF and laser-induced), flat-round transform, crab cavities, 2-frequency RF, etc.)?
9. Can longitudinal emittance be reduced?
10. What is the optimal RF frequency or combination of frequencies?
11. Can lasing be achieved (SASE FEL, X-ray FEL oscillator)?
12. Do tracking and simulation codes need development? How can these codes be built to match real machines to ensure achieving the predicted performance?
13. What studies can be performed on existing storage rings?
14. What R&D is needed before an actual USR is built?

Not all questions were addressed in the Workshop, but they remain for future consideration. Some talks in other sessions, such as Lattice Design and IDs, are also accelerator physics related. These talks addressed the issues of round beam, ID effects, etc.

Presentations given in this session are listed in Appendix A.2. The following items were noted in the session and discussion periods:

**Modeling and simulation:**

1. Codes for collective effects for USRs are in various stages of completeness (rated 1-5, 5 highly complete): Touschek lifetime (5), IBS (4), impedance (3), ion instability (2), CSR (2), space charge for low-E rings (1)).
2. Codes/formulas should be benchmarked on working machines that can approximate USR parameters by reducing energy, coupling, etc. (e.g. PETRA-III, ESRF, SPring-8).
3. General scaling laws that take into account as much as possible all the effects, including emittance (with collective effects), brightness, spectrum, circumference, magnet strengths, running costs, etc.).

**Collective effects, impedance and lifetime:**

1. IBS, collective effects and Touschek lifetime are serious but not insurmountable issues that affect lattice and hardware design.
2. The contribution to emittance growth from IBS for 1-2-km circumference rings is minimized when electron energy is between ~4.5 and 6 GeV.
3. Ways to suppress CSR from short bunches were not discussed.
4. Ion instabilities were not discussed.

**Dynamic aperture:** see comment on ESRF dynamic aperture optimization in the Lattice Design session, section 3.1.

**RF:**

1. Need general scaling laws that take into account as much as possible all the effects, including energy acceptance, bunch length, RF power, equipment size, costs, etc.
2. Higher frequency SC RF (~1.5 GHz) should be considered for generating short bunches with high peak current; 2 RF frequencies operating in beat frequency mode (e.g. x3 and x3.5 of base 500 MHz) should be considered for generating long/short bunches.
3. SC vs. NC: best solutions are subject to existing infrastructure and expertise.
4. See Accelerator Engineering section for more on frequency optimization.

**Short bunches:**

1. High rep-rate short bunches (~ps or less) are of interest for users; >$10^6$ ph/pulse desired.
2. Production methods include low $\alpha$, harmonic cavities, crab kickers, $\varepsilon$ exchange, etc
3. <0.1 ps bunch propagation limited to very few turns by CSR (increasing length, $\varepsilon$).

**Lasing:**

1. Initial studies show that single-pass nm lasing in a switched bypass may be possible using a vertically oriented transverse gradient undulator (TGU) and pm vertical emittance if <200 Apk can be achieved. Oscillator configurations may also be possible
2. Ways to achieve 200 Apk with $\varepsilon$~10-pm and/or to reduce energy spread not discussed.
3. Localized compression should be considered for high peak current FEL bunches to avoid HOM heating issues in ring.

**Reduced energy spread, longitudinal emittance:**

1. Related to the above, explore ways to reduce energy spread and longitudinal emittance in general, to enable using high ID harmonics, short bunches and potential lasing. High frequency, high voltage RF (~1.5 GHz) may be one option.

**IDs:**

1. ID effects on orbit, optics, dynamic aperture, energy spread and impedance are all enhanced in USRs and impact accelerator engineering.

Members of the Accelerator Physics Working Group included K. Bane, M. Borland, M. Boscolo, D.H. Ji, E. Levichev, Q. Qin (co-chair), P. Raimondi (co-chair), K. Soutome, S.K. Tian, J.Q. Wang, and N. Wang.

### 3.3 Injection

Questions for consideration in these sessions included:

1. What are injection orbit transient requirements for USRs and how can they be achieved? This depends on kicker pulse length, ring revolution period, etc.
2. What are best off-axis injection systems for frequent top-up injection?
3. Can on-axis injection satisfy top-up current constancy needs for low-lifetime USRs?
4. What are on-axis injection options and associated injector requirements (linac injector, booster, accumulator ring, accumulator/booster, number of bunches per injection, etc.)
5. What are septum, kicker, linac performance requirements for top-up injection into USRs?
5. Do tracking and simulation codes need development?
6. What studies can be performed on existing storage rings?
7. What R&D is needed before an actual USR is built?

Not all questions were addressed in the Workshop, but they remain for future consideration.

Presentations given in this session are listed in Appendix A.2. The following items were noted in the session and discussion periods:

**Top-Up Injection:**

1. Sensitivity to residual orbit transient will be much greater in USRs.
2. Multi-shot injection allows small topping-up of several arbitrary bunches, but extends orbit transient time. **Ideal:** arbitrary bunch pattern top-up in one shot.
3. Pulsed multipole injection schemes are being developed to replace traditional kicker bump injection and reduce residual orbit transient.
4. Low injected beam emittance (a function of USR acceptance, dynamic aperture) needed for good injection efficiency. Booster in ring tunnel? Linac?

**Pulsed Multipole Injection:**

1. Several pulsed multipole (PM) injection schemes in development (quadrupole, sextupole, "nonlinear" kickers, TEM-mode kickers). **Ideal:** septum and pulsed multipole in same straight (e.g. Sirius)
2. SPring-8 considering a scheme using a pulsed quadrupole together with another upstream pulsed quad to suppress quad mode oscillation.
3. Need low injected beam emittance for high Injection efficiency, otherwise injected beam samples different kicks in PM.

4. If PM injection is chosen for new rings, then it should be included as a design requirement for the injection area configuration.

**On-axis Injection:**

1. On-axis "swap-out" injection enables injection into low-dynamic aperture USRs; single-bunch and bunch-train injection schemes proposed.

2. For single-bunch or short bunch-train injection, need fast rise/fall time kickers (<2 ns rise and fall, otherwise need gap between bunches); for long bunch-train injection, need very flat top kickers (~$10^{-3}$ of kick amplitude)

3. Linac or linac + booster injector limits total charge that can be swapped out and thus total current in ring (< ~200 mA for km rings)

4. Swap-out injection for large rings is best served with an accumulator ring, adding large cost unless the accumulator can also be a booster.

5. A scheme to recover swapped-out beam in an accumulator for re-injection into USR after topping up has been suggested (Borland); orbit transients may be an issue.

**Injection magnets and kickers:**

1. SPring-8 designing fast (4 ns) TEM-mode injection kickers that can be driven either as a quadrupole for off-axis injection (2-mm offset) or a dipole for on-axis injection.

2. BESSY stripline nonlinear kicker burned up: modified design to hide conductors. ILC kicker pulsers have <2 ns rise time, <2 ns flat-top, but fall time is longer with overshoot.

3. Shrinking structures increases machining and alignment requirements; field mapping very difficult for small aperture structures.

Members of the Injection Working Group included J.H. Chen, O. Dressler, R. Hettel (co-chair), Y. Jiao (co-chair), A. Kling, S. Leemann, L. Liu, K. Soutome and W. Kang.

## 3.4 Accelerator Engineering

Questions for consideration in these sessions included:

1. Can the high gradient magnets proposed for various USR designs be built in a practical way? Are superconducting magnets needed? What are the limits to high gradient magnets?

2. Can there be a leap in combined function multipole magnet technology?

3. What are the vacuum chamber aperture requirements for various USR designs and can the chambers be built in a practical way? What are the aperture limits?

4. What are the injection kicker requirements for various USR designs (e.g. rise/fall times, flat-top constancy, transverse field constancy) and can they be achieved?

5. What level of injection charge constancy can be achieved?

6. What are the mechanical alignment and stability requirements for various USR designs and can they be achieved in a practical way? What are the achievable limits for alignment and stability? What are the ground stability requirements?

7. Are there any RF system design issues?

8. How can ring power consumption be minimized? (very important for future rings)

9. What studies can be performed on existing storage rings?

10. What engineering R&D is required for USR implementations?

Not all questions were addressed in the Workshop, but they remain for future consideration.

Presentations given in this session are listed in Appendix A.2. The following items were noted in the session and discussion periods:

**General:**

1. Technologies for magnets, vacuum, RF and stability are strongly interconnected. A high degree of system integration is necessary.

2. USRs consist of a very large number of magnet elements, some having strong multipole gradients $\Rightarrow$ small magnet bores, small aperture chambers with distributed pumping, special designs for extracting light, sub-micron stability requirements.

**Magnets:**

1. Very strong multipole magnets not a big problem as long as bores are small, but emphasis should be put on reducing gradients in the lattice design process.

2. Mechanical tolerances for small-bore magnets are very strict and ultimately limit bore radius. Over-specification of tolerances can lead to high production costs.

3. Stability and alignment issues should be addressed in the magnet design. Options include machining several magnets in the same iron block (MAX IV), which can have lowest vibration eigenfrequencies in the 100-Hz range.

4. Operating magnets in non-linear part of excitation curve is sometimes needed. R&D on material selection and permanent magnet solutions is needed.

**Injection magnets and kickers:** see Injection session notes.

**Vacuum system:**

1. Constraints to reducing vacuum pipe bore include injection needs, NEG coating requirements and mechanical tolerances. The likely minimum is ~ 8 mm.

2. NEG-coating requires complex and delicate process of etching, cleaning and coating. R&D is needed to investigate procedures for coating small-bore chambers, minimizing coating and activation times, and time needed for intervention procedures.

3. Industrial NEG-coating capacity is presently a bottle-neck.

4. The impedance of the small-bore chambers, especially with short bunches, is an issue.

5. Distributed heat absorbers, using the vacuum chamber itself, should be considered. R&D is needed to investigate the choice of materials for vacuum chambers.

**RF system:**

1. Low frequency systems providing longer bunches offer advantages for medium energy USRs. Longer bunches, made even longer with harmonic cavities, offer passive stability for many collective effects, low electricity consumption and low investment cost.

2. For short bunches, or if the energy losses/turn exceeds a critical value (~ 2-3 MeV/turn), low-freq RF is bulky and the high shunt impedance of higher freq systems is preferred.

3. As emittance is reduced below 100 pm and the demand for short bunches and/or high peak currents are increased, ring energy will likely increase to overcome IBS and RF systems in the GHz region (probably SC) may be preferred.

**Stability:**

1. USR beam dimensions are very small, resulting in sub-micron stability requirements. Girder vibration and lattice amplifications will decrease displacement tolerances. FOFB and other component feedback systems will help suppress beam motion.

2. When choosing a USR construction site, a careful study of the ground geological composition and its properties should be carried out. As with 3$^{rd}$ generation rings, FEA calculations including buildings should also be carried out and strict rules for locating machinery should be established.

3. The accelerator and beam line floors and buildings should be optimized so as not to amplify the vibrations. Vibrations damping should be considered.

**Not discussed:**

1. Bunch compression, emittance exchange, other bunch manipulation systems using RF and/or lasers.

2. Beam cooling systems.

3. Magnet power supplies.

4. Advanced alignment methods.

5. High repetition rate kickers for bypass switching.

6. Ways to reduce power consumption.

7. Value engineering.

Members of the Accelerator Engineering Working Group included R. Bartolini, F.S. Chen, E. Al-Dmour, M. Eriksson (co-chair), M. Johansson, G. Kulipanov, G.H. Luo, H.M. Qu (co-chair), L. Rivikin, C. Zhang and L. Zhang.

## 3.5 Instrumentation and Feedback Systems

Questions for consideration in this session included:

1. What are BPM and feedback system requirements needed to achieve sufficient stability?
2. What are photon BPM requirements and associated photon beam line design requirements?
3. How can higher order lattice parameters be measured and corrected (e.g. higher order resonance cancellation, etc.)?
4. What are the best monitors for bunch dimensions (transverse and longitudinal) and coherence?
5. What studies can be performed on existing storage rings?
6. What R&D is needed before an actual USR is built?

Not all questions were addressed in the Workshop, but they remain for future consideration.

Presentations given in this session are listed in Appendix A.2. The following items were noted in the session and discussion periods:

**General:**

1. 3rd generation light sources are already operating at the diffraction limit in the vertical plane and suitable beam instrumentation and feedback technology largely exists.
2. Nevertheless, the production of very low emttance and highly coherent x-rays and their faithful transport to experimental stations in USR facilities will likely require improved performance and the integration of ring and beam line x-ray stabilizing systems.

**e- BPMs:**

1. New designs for BPM pick-ups in small aperture vacuum chambers should be considered, providing sufficient sensitivity and miniimal impedance.
2. New designs for shielded bellows for small aperture BPM should be considered. Bellows on both sides of a BPM is better than on just one side.
3. BPMs should be referenced to quadrupoles or sextupoles in arcs; referenced to ground in straight sections. Invar is an excellent support material given any thermal fluctuations, but other materials are suitable if temperature is highly stable.
4. Modern BPM processing systems already match most USR requirements, but improving turn-turn resolution by x10 would enable measuring and correcting higher order resonance driving terms to improve dynamic aperture – a critical capability. Improvements to current-dependence and latency time are also desirable.

**X-ray BPMs:**

1. Performance of photon monitors (X-BPMs) for hard x-ray planar undulators is reasonable. "Decker distortions" in straight sections can help to improve X-BPM performance by using low-field dipoles at the straight section entrances and exits to redirect unwanted radiation from much stronger upstream dipoles.

2. No good photon monitor solution yet for VUV ID and soft x-ray EPU radiation.

3. One potential approach for solving difficult photon position monitoring problems is to deduce position based on information from beam line detectors and other diagnostics.

**Beam size monitors and stability:**

1. Micron level resolution is required for transverse beam size measurement, especially for horizontal plane.

2. The following measurement methods should be evaluated carefully to make sure qualified for USRs-----X ray: FZP, CR Lens, interferometry, K-B mirror, B-F Lens.

3. Beam size stability is an important issue for light sources – feed-forward and feedback systems in use at many rings.

4. Need to simulate sensitivity of possible round beam schemes (coupling, mobius, wigglers, dispersion) to errors and develop stabilization strategy.

**Orbit feedback:**

1. No revolution in orbit feedback is necessary, but continued development to improve performance is needed.

2. Integration of improved photon BPMs, other beam line diagnostics, hydrostatic level or equivalent sensors, possibly beam line detector information, etc. into a unified beam stabilizing feedback system is envisioned as a way to maximize beam stability.

3. BPM/feedback update rates have improved (1 kHz -10 kHz). Latency times of digital BPMs are typically more than one 10-kHz cycle and need to be reduced.

4. Hydrostatic leveling or equivalent sensors to monitor the motion of accelerator and beam line components with <200-nm resolution need to be developed.

**Mulitbunch feedback (MBFB):**

1. High resolution BPM (< 1 um) with narrow beam pipe is required for USR MBFB.

2. Higher gain, better ADCs (> 12 bits) and more sensitive pickups are required.

3. MBFB operation with "hybrid" filling patterns, where one or more bunches has substantially more charge than the others, needs attention (e.g. a bunch current-sensitive front end attenuator may be needed for the system).

4. The effect of MBFB noise on beam size needs evaluation and mitigation as needed.

Members of the Instrumentation and Feedback Working Group included J. Cao (co-chair), J.-C. Denard, T. Fujita, S. Kurokawa, Y.B. Leng, C. Steier, J.H. Yue and Z.T. Zhao (co-chair).

## 3.6 Insertion Devices

Questions for consideration in this session included:

1. What undulator parameters are needed to best exploit diffraction-limited beam emittance (gap, period, phase error, etc.)
2. What performance can be expected from future superconducting IDs?
3. Are there novel ID structures to be developed for unique applications?
4. Are there ID structures to be developed to reduce power on optics?
5. What are damping wiggler parameter requirements for 10-pm rings?
6. Can transverse gradient undulators be used for ring-based FELs?
7. Is there a role for fast-switching or pulsed RF undulators?
8. What studies can be performed on existing storage rings?
9. What R&D is needed before an actual USR is built?

Not all questions were addressed in the Workshop, but they remain for future consideration.

Presentations given in this session are listed in Appendix A.2. The following items were noted in the session and discussion periods:

**General:**

1. Conventional planar IDs appropriate for many users, particularly for higher energy rings.
2. Differing views are taken on the use of damping wigglers for emittance reduction by those considering upgrades to existing machines as opposed to those considering green-field proposals. Those considering upgrades of existing machines, where the ability to reduce emittance by lattice alone is constrained by existing circumference and beam line locations, appear to be more likely to consider using damping wigglers than green field ring designers. These considerations are reflected in issues below.
3. Ongoing R&D on CPMUs, SCUs, variable polarizable and other new IDs will benefit USRs.
4. USRs may enable smaller ID gaps, limited by impedance effects and heating.
5. Vertically oriented and small-bore helical IDs might be accommodated in USRs.

**Issues:**

1. ID changes may impact USR performance in a greater way. Need precision compensation of tune, beta beat, beam size, emittance, orbit and dynamic effects.
2. Energy spread will be an important issue, impacting higher harmonics.
3. Heat load and power density may be issues, especially for high energy machines.
4. Damping wigglers produce large amounts of synchrotron radiation on accelerator components, increasing the risk of component damage.

5. The lengths of IDs and straight sections need careful optimization, an issue that influences the electron/photon emittance-matching issue discussed above and brings into question the pros and cons of placing two IDs in a straight section in either canted or collinear configuration as a way to increase the number of x-ray beam ports or enhance beam line performance.

6. Magnetic environments in straight sections need special attention: background field, magnetic materials in ID, magnet fringe fields, ion pumps, etc.).

7. ID error tolerances need to be defined by lattice designers and ID users.

8. ID commissioning in USRs will be more complex, and operation will require improved x-ray beam instrumentation.

9. X-ray optical components will need improvement to realize improved beam qualities from USR IDs.

10. New techniques for measuring the field quality of future insertion devices are needed.

Members of the Insertion Device Working Group included J. Bahrdt, J. Chavanne, R. Gerig (co-chair), M. Jaski, M. Li, H.H. Lu, Y.Z. Wu (co-chair), L.X. Yin, Q.G. Zhou and K. Zolotarev.

# 4. Summary of Accelerator R&D Topics for USRs

The following R&D topics were identified in the Workshop sessions, presented without any attempt at prioritization.

## 4.1 Lattice Design

**Low emittance, buildable lattices:** Develop low emittance lattice designs having "reasonable" multipole gradients and magnet apertures. Explore benefit of using dipoles with longitudinal gradient.

**Design optimization:** Optimize ring parameters (e.g. energy, emittance, circumference, beta functions, RF, etc.) based on targeted spectral brightness, coherence, special operating modes (e.g. short bunches, lasing) and number of beam lines. Define a quality factor to gauge this optimization. Develop optimization algorithms. Present "envelope of performance" showing trade-offs in emittance, beam current and bunch length.

**Consolidated beam lines:** Develop lattice geometries, potentially non-circular and/or having hybrid lattices and/or straight sections that hold more than one ID, that enable consolidating beam line straight sections in very large rings – a part of design optimization.

**Robinson wigglers**: Are they a replacement for conventional damping wigglers in reducing emittance?

**Round beams:** Determine optimal ways to produce nearly round beams at source points. Test on existing machines if possible.

**Momentum compaction:** Develop very low emittance lattices with increased momentum compaction as a way to increase bunch length (e.g. chicanes, etc.).

## 4.2 Accelerator Physics

**Simulation codes:** Develop codes that account for close magnet spacing and include collective effects during lattice optimization. Improve simulation codes impedance, ion instability, CSR and other effects as needed. Benchmark and calibrate codes on existing machines operating in special modes that approximate USR operating conditions.

**Scaling laws:** Develop general scaling laws that take into account as much as possible all the effects, including emittance (with collective effects), brightness, spectrum, circumference, magnet strengths, RF, running costs, etc.

**Non-linear lattice correction**: Develop improved techniques to measure and correct higher order resonance driving terms to maximize dynamic apertures. Test on existing rings.

**Short bunches and RF frequency:** Study the benefit of higher RF frequency for reducing longitudinal emittance, bunch length and operating costs, and the use of using 2 frequencies to generate alternating long and short bunches.

**Reduced energy spread, longitudinal emittance:** Explore ways to reduce energy spread and longitiudinal emittance in general, to enable using high ID harmonics, short bunches and potential lasing.

**Very short bunches and CSR:** Explore ways to suppress CSR to reduce the lengthening and emittance increase of very short bunches propagating in the ring.

**Ion instabilities:** Determine how ion instabilities can be mitigated.

**High peak current:** Explore ways to produce >200 Apk with 10 pm-scale emittance to enable lasing.

**Lasing:** Determine beam parameters and consequent ring designs that would enable X-ray FEL operation, either in a switched bypass or in the ring itself, including oscillator configurations.

**Beam manipulation**: Explore ways to (locally) reduce emittance, bunch length, energy spread, etc. (e.g. emittance exchange, flat-to-round converter (ID in solenoid), RF and optical manipulation methods, etc.).

**Space charge:** Determine if space charge is an issue for low-E USRs.

## 4.3 Injection

**Single-shot top-up:** Ways to restore charge to multiple arbitrary bunches in a single injection shot to reduce the duration of the top-up-related orbit transient, maintaining variation in charge for all bunches to ~20% or less for a uniform fill pattern.

**Pulsed multipole (PM) injection:** Continued development of PM injection schemes, including schemes with septum and PM in the same straight.

**Accumulator/booster for swap-out injection:** Study the practicality of implementing a combined accumulator/booster, possibly located in the main ring tunnel, for realizing multibunch single-shot swap-out injection. Investigate the possibility of recovering the beam kicked out from the ring in the accumulator/booster for reinjection.

**Injection kickers:** See Accelerator Engineering.

**Longitudinal injection**: Investigate practicality of longitudinal injection as a way to eliminated stored beam orbit transient.

## 4.4 Accelerator Engineering

**Magnets:** Determine optimal magnet bore dimensions with respect to mechanical tolerances, multipole strengths, yoke saturation and vacuum system design. Investigate magnet material choice, solid versus laminated cores and compact combined function magnet designs.

**Vacuum system:** Designs for small aperture vacuum systems with focus on chamber material, NEG coating and activation processes, heat absorption, synchrotron light extraction and BPM head stability.

**Stability**: Develop site vibration specifications for USRs. Develop passive and active ways to minimize effects on the stability of the photon beam and critical accelerator and beam line components caused by ground motion, cooling water, machine- and temperature-induced motion and vibration. Develop stable building design concepts.

**Motion sensors:** Develop affordable 100-nm-resolution component motion sensors.

**Alignment:** Develop practical and simplified ways to achieve 10-μm alignment tolerances.

**RF system:** Optimal frequency(s), improved cavity mode damping, solid state amplifiers, harmonic cavity systems (including passive vs. active), crab and other beam manipulation cavities, solid state RF power sources, continued improvements to LLRF.

**Power supplies:** Not discussed.

**Pulsed multipole injection magnets:** Designs that reduce the required separation of injected and stored beams.

**Fast kickers:** Develop injection kicker and pulser designs having <4 ns total baseline pulse width for swap-out injection of single bunches separated by 2 ns.

**Flat-top kickers:** Develop long-pulse injection kicker and pulser designs that have flat-top constancy on the order of $10^{-3}$ of full amplitude over the order of 100 ms for multi-bunch swap-out injection.

**High repetition rate kickers:** 10-100 kHz, fast rise/fall times for deflecting beam into bypass or other beam manipulation.

**Field mapping:** Field mapping devices and techniques for small aperture magnets, kickers.

**Power consumption:** Ways to reduce accelerator power consumption.

## 4.5 Instrumentation and Feedback Systems

**e- BPMs:** Stable BPM designs for small aperture vacuum chambers having micron turn-turn resolution or better.

**BPM processors:** A factor of 10 or more increase in turn-turn resolution than present state-of-the-art for measurement of higher order lattice resonance driving terms; reduced processing latency to be commensurate with 10-kHz digital feedback clock rates; improved stability and reduced current dependence.

**X-ray BPMs:** Continued development of photon BPMs for IDs, especially EPUs and VUV. Photon BPMs located close to experimental sample (e.g. 4-quadrant thin crystal scatterer, etc.)

**Orbit feedback:** Integrated orbit and beam line component feedback systems to achieve maximal beam stability at the experiment using multiple sensor types (e.g. e-BPMs, X-BPMs, beam line sensor and detector information, motion monitors, etc).

**Beam size stabilization:** Feedback and feedforward systems to stabilize beam size as IDs, especially EPUs, are varied.

**Multibunch feedback:** Improved systems having higher resolution, reduced noise impact and capable of accommodating variable bunch fill patterns, including ones with single large bunches and many small ones.

## 4.6 Insertion Devices

**New IDs:** Continue ongoing R&D on CPMUs, SCUs, variable polarization and other new IDs will benefit USRs.

**ID length:** Establish optimal lengths for IDs in USRs; straight section lengths should be determined accordingly.

**Small gaps:** Determine minimum ID gaps.

**Vertically oriented IDs:** Can they be accommodated (e.g. Delta-type, helical, TGUs, etc.)?

**Power on optics:** Develop improved masking schemes and IDs that minimize unused power on optics.

**Dynamic effects:** Establish ID tolerance requirements and study effects of present and anticipated future IDs and USR beam dynamics and properties and develop effective compensation schemes.

**ID commissioning:** Develop new ID commissioning strategies as needed for USRs; test on existing machines.

**X-ray optics:** Develop X-ray optical components capable of preserving photon beam properties, including coherence, from USR IDs.

**Modeling codes:** Develop codes for the generation of X-rays in IDs and their wavefront propagation in photon beam lines that accurately account for possibly complicated ID structures, varying electron parameters within IDs, etc.

# 5. Next Steps

The Workshop on Accelerator R&D for Ultimate Storage Rings may have helped to more clearly define USR accelerator design issues and the areas where R&D are needed. On the other hand, many issues were not addressed at the workshop, and many new questions came to the surface. The USR design process is an ongoing effort which requires much more future work by the storage ring light source community. In the near term, the following steps are suggested for consideration:

**Definition of the science case:**

- The science case of USRs needs to be more clearly defined so that facility designs can be better optimized for cost and benefit.
- A series of international workshops has begun to define the science case; these will hopefully continue with the possibility of developing the scientific justification for 10-pm-scale (or less) storage rings.

**Continued accelerator workshops:**

- Future workshops on more focused USR accelerator topics are needed.
- The integration of USR workshops with other low emittance ring workshops (e.g. LowEring) would be beneficial.
- The formation of ongoing working groups for various topics would be beneficial.

**Definition of R&D for beam line and optics design:**

- Technical challenges for USR x-ray beam line and optics designs are formidable, including power absorption, coherent wavefront preservation, micro-manipulation techniques, component stabilization, high resolution/high rep rate detectors, etc.
- An R&D program needs to be defined, perhaps using workshops.

**Support for USR R&D program:**

- While individual light source facilities may have the ability to fund some level of R&D for USR design using lab R&D or operation funds, a sustained R&D program lasting several years, especially if it involves developing hardware components, will likely require special support from the national funding agencies.
- A comprehensive USR R&D plan that includes scope, budget and schedule should be developed in preparation for seeking funding from national agencies. Collaboration between institutions, both national and international, could help eliminate duplication of efforts and strengthen the case for funding.

# References

[1] M. Bei, M. Borland, Y. Cai, P. Elleaume, R. Gerig, K. Harkay, L. Emery, A. Hutton, R. Hettel, R. Nagaoka, D. Robin, C. Steier, "The Potential of an Ultimate Storage Ring for Future Light Sources", Nucl. Instr. and Meth. Phys. Research A, 622 (2009) 518-535.

[2] http://www-conf.slac.stanford.edu/icfa2010/Proceedings.asp

[3] http://www.jlab.org/conferences/FLS2012/

[4] http://erl.chess.cornell.edu/gatherings/2011_Workshops/

[5] http://lowering2011.web.cern.ch/lowering2011/

[6] D. Einfeld, M. Plesko, NIM-A 335, p.402-416, 1993.

[7] M. Borland, "Progress Towards Ultimate Storage Ring Light Sources", Proc. IPAC 12, New Orleans, 2012.

[8] A. B. Temnykh, "Delta Undulator for Cornell Energy Recovery Linac", Phys. Rev. Special Topics - Accelerators and Beams. 2008; 11:120702.

[9] D. Robin, ICFA FLS 2010 (http://www-conf.slac.stanford.edu/icfa2010/Proceedings.asp)

[10] Z. Huang, Y. Ding, C. Schroeder, "Compact X-ray Free-Electron Laser from a Laser-Plasma Accelerator Using a Transverse-Gradient Undulator", PRL 109, 204801 (2012).

[11] K.-J. Kim, internal technical note, 10/9/12.

**Appendix A.1:  Workshop Poster**

# Workshop on
## Accelerator R&D for Ultimate Storage Rings
### HUAIROU - BEIJING, Oct. 30 – Nov. 1, 2012

To review worldwide efforts in designing ultimate storage rings (USRs) and to identify technical challenges requiring research and development.

- Overview of USR design
- Parameter optimization
- Lattice design
- Collective effects
- Injection issues
- Beam stability
- Engineering challenges

**Workshop Co-Chairs**
Q. Qin — IHEP
R. Hettel — SLAC

**International Advisory Committee**
K. Balewski — DESY
A.W. Chao — SLAC
H. Ding — IOP
M. Eriksson — MAX-Lab
R. Gerig — APS
X.M. Jiang — IHEP
H. Ohkuma — SPring-8
P. Raimondi — ESRF
Z. Zhao — SINAP

**Local Committee**
Q. Qin — IHEP
G. Xu — IHEP
Y. Jiao — IHEP
Q. Pan — IHEP
M.J. Yu — IHEP
J. Zhou — IHEP

**Scientific Program Committee**
R. Bartolini — DLS
M. Borland — ANL
Y. Cai — SLAC
J. Chavanne — ESRF
M. Johansson — MAX-Lab
S. Krinsky — BNL
G. Kube — DESY
E. Levichev — BINP
A. Nadji — SOLEIL
K. Soutome — Spring-8
C. Steier — LBNL
A. Streun — PSI
J. Wang — IHEP
L. Yin — SINAP

Contact: jiaoyi@ihep.ac.cn
Workshop Website: http://usr2012.ihep.ac.cn/

*Hosted by Institute of High Energy Physics, Chinese Academy of Sciences*

**Appendix A.2: Workshop Agenda**

# Schedule of the Workshop on Accelerator R&D for USRs

## (Huairou, Beijing, China, Oct. 30 – Nov. 1, 2012)

**Oct. 29**
    Registration      from 12:00 pm
    Reception       19:00 pm

**Oct. 30**

| | **Welcome and Introduction Session** | | |
|---|---|---|---|
| 08:30 – 08:40 | Welcome | Yifang Wang (IHEP DG) | Chair: X.M. Jiang |
| 08:40 – 08:50 | Workshop charge | Qing Qin (IHEP) | |
| 08:50 – 09:05 | Scientific cases for USR | Hong Ding (IoP) | |
| 09:05 – 09:20 | Highlights of science of USR | Yuhui Dong (IHEP) | |
| 09:20 – 09:35 | Overview of USR design and performance requirements | Robert Hettel (SLAC) | |
| | **Session 1:    Lattice Design (part 1)** | | |
| 09:35 – 09:55 | MAX-IV project and MBA with higher multipoles | S. Leemann (Max-Lab) | Chair: G. Xu |
| 09:55 – 10:15 | Latest development of ESRF and future plan | P. Raimondi (ESRF) | |
| 10:15 – 10:35 | Lattice design for Spring-8 II | Y. Shimosaki (SPring-8) | |
| 10:35 – 10:50 | Group photo and Coffee break | | |
| | **Session 2:    Lattice Design (part 2)** | | |
| 10:50 – 11:07 | USR upgrade for ALS | C. Steier (LBNL) | Chair: H. Ohkuma |
| 11:07 – 11:24 | Study of lower emittance lattice at SOLEIL | A. Nadji (SOLEIL) | |
| 11:24 – 11:44 | BAPS lattice design | G. Xu (IHEP) | |
| 11:44 – 12:15 | Discussion | | |
| 12:15 – 13:30 | Lunch | | |
| | **Session 3:    Accelerator Physics (part 1)** | | |
| 13:30 – 13:50 | Collective effects in USRs | K. Bane (SLAC) | Chair: P. Raimondi |
| 13:50 – 14:10 | Single and multiple Touschek scattering in low emittance rings | M. Boscolo (INFN-LNF) | |
| 14:10 – 14:30 | BAPS CSR, impedance and IBS | N. Wang/S.K. Tian (IHEP) | |
| 14:30 – 14:50 | Propagation of ultra-short bunches in USRs and potential lasing | X.B. Huang (SLAC) | |
| 14:50 – 15:05 | Optimal RF frequency | M. Eriksson (MAX-Lab) | |
| 15:05 – 15:30 | Discussion | | |
| 15:30 – 15:45 | Coffee break | | |
| | **Session 4:    Accelerator Engineering (part 1)** | | |
| 15:45 – 16:05 | Some engineering challenges for a USR machine | L. Zhang (ESRF) | Chair: M. Eriksson |

| 16:05 – 16:25 | Anticipated limits on machine stability and consequences on machine and beamline designs | J.C. Denard (SOLEIL) | |
| --- | --- | --- | --- |
| 16:25 – 16:45 | MAX-IV magnet design | M. Johansson (MAX-Lab) | |
| 16:45 – 17:00 | High Gradient Magnet Design | K. Soutome (SPring-8) | |
| 17:00 – 17:15 | Design and fabrication of NSLS-II sextupole | W. Kang (IHEP) | |
| 17:15 – 17:30 | Design for high precision small aperture magnets for BAPS | F.S. Chen (IHEP) | |
| 17:30 – 18:00 | Discussion | | |
| 18:30 | Dinner | | |

## Oct. 31

| | Session 1: Lattice Design (part 3) | | |
| --- | --- | --- | --- |
| 08:00 – 08:20 | Compact low emittance lattice including superbends | L. Liu (LNLS) | Chair: J.Q. Wang |
| 08:20 – 08:40 | PEP-X design | J. Safranek (SLAC) | |
| 08:40 – 09:00 | Tevatron USR and other topics of USR | M. Borland (ANL) | |
| 09:00 – 09:20 | Overview of optics correction at Diamond and plan for DLS-II | R. Bartolini (DLS) | |
| 09:20 – 09:37 | Round beam experience at BINP | E. Levichev (BINP) | |
| 09:37 – 10:15 | Discussion | | |
| 10:15 – 10:30 | Coffee break | | |
| | Session 2: Accelerator Physics (part 2) | | |
| 10:30 – 10:50 | Optimization of dynamic aperture for the ESRF upgrade | A. Franchi (ESRF) | Chair: Q. Qin |
| 10:50 – 11:10 | Coupling correction at SLS, minimum vertical emittance | L. Rivkin (PSI) | |
| 11:10 – 11:25 | 2-frequency RF system for short bunches at SPring-8 | T. Fujita (SPring-8) | |
| 11:25 – 12:00 | Discussion | | |
| 12:00 – 13:30 | Lunch | | |
| | Session 3: Injection (part 1) | | |
| 13:30 – 13:50 | Top – up injection | L. Rivkin (PSI) | Chair: Z.T. Zhao |
| 13:50 – 14:10 | Injection into low-lifetime USRs | R. Hettel (SLAC) | |
| 14:10 – 14:30 | Pulsed sextupole injection for MAX-IV | S. Leemann (MAX-Lab) | |
| | Session 4: Injection (part 2) | | |
| 14:30 – 14:50 | Pulsed sextupole injection for BAPS | Y. Jiao (IHEP) | Chair: R. Hettel |
| 14:50 – 15:10 | Injection scheme for the SPring-8 upgrade | K. Soutome (SPring-8) | |
| 15:10 – 15:30 | On-axis swap-out injection | M. Borland (ANL) | |
| 15:30 – 16:00 | Discussion | | |
| 16:00 – 16:15 | Coffee break | | |
| | Session 5: Accelerator Engineering (part 2) | | |
| 16:15 – 16:35 | MAX-IV vacuum chamber design | Eshraq Al-Dmour (MAX-Lab) | Chair: H.M. Qu |

| 16:35 – 16:55 | Single non-linear injection kicker magnet | O. Dressler (BESY-II) | |
| 16:55 – 17:15 | Fast pulsed power supply for ILC damping ring kicker | J.H. Chen (IHEP) | |
| 17:15 – 17:30 | Ground vibration for BAPS | D.H. Ji (IHEP) | |
| 17:30 – 18:00 | Discussion | | |
| 18:30 – 20:00 | Banquet | | |

**Nov. 1**

| | | | |
|---|---|---|---|
| **Session 1: Instrumentation and Feedback** | | | |
| 08:00 – 08:25 | SOLEIL orbit feedback systems and photon BPMs | J.C. Denard (SOLEIL) | Chair: J.S. Cao |
| 08:25 – 08:45 | Multi-bunch feedback system in USR | T. Fujita (Spring-8) | |
| 08:45 – 09:05 | Feedback system for USRs | C. Steier (ALS) | |
| 09:05 – 09:35 | Discussion | | |
| **Session 2: Insertion Devices** | | | |
| 09:35 – 09:55 | IDs for the ESRF upgrade | J. Chavanne (ESRF) | Chair: R. Gerig |
| 09:55 – 10:15 | Superconducting IDs at BINP | K. Zolotarev (BINP) | |
| 10:15 – 10:30 | Coffee break | | |
| 10:30 – 10:50 | IDs for Shanghai Synchrotron Radiation Facility | Q.G. Zhou (SINAP) | |
| 10:50 – 11:10 | PETRA-III damping wiggler experience | A. Kling (DESY) | |
| 11:10 – 11:30 | Undulator technology for USR | M. Jaski (ANL) | |
| 11:30 – 11:50 | Cryogenic permanent magnet and superconducting small period undulators | J. Bahrdt (BESSY-II) | |
| 11:50 – 12:30 | Discussion | | |
| 12:30 – 13:30 | Lunch | | |
| **Closing Session** | | | |
| 13:30 – 15:00 | Report writing | | |
| 15:00 – 15:15 | Coffee break | | |
| 15:15 – 16:00 | Summary | | |
| 16:00 | Closeout and Departure | | |

# Appendix A.3:  Workshop Attendees

**ANL/APS:**  M. Borland, R. Gerig, M. Jaski

**BINP:**  G. Kulipanov, E. Levichev, K. Zolotarev

**Cosylab:**  S. Kurokawa

**DESY:**  A. Kling

**Diamond:**  R. Bartolini

**ESRF:**  J. Chavanne, A. Franchi, P. Raimondi, L. Zhang

**Helmholtz/BESSY-II:**  J. Bahrdt, O. Dressler

**IHEP:**  J.S. Cao, F.S. Chen, J.H. Chen, S.Y. Chen, Y.H. Dong, D.H. Ji, X.M. Jiang,  Y. Jiao, W. Kang, M. Li, H.H. Lu, Q. Pan, Q. Qin, H.M. Qu, S. Tian, J.Q. Wang, N. Wang, S.H. Wang, Y.F. Wang, L. Wu, Y.Z. Wu, G. Xu, M.J. Yu, J.H. Yue, C. Zhang, N. Zhao

**INFN-LNF:**  M. Boscolo

**Institute of Physics (Beijing):**  H. Ding

**Kurchatov Institute:**  V.N. Korchuganov (absent)

**LBNL/ALS:**  C. Steier

**LNLS:**  R. Farias, L. Liu

**MAX-Lab**:  E. Al-Dmour, M. Eriksson, M. Johansson, S. Leemann

**NSRRC:**  G.H. Luo

**PSI:**  L. Rivkin

**SINAP:**  Y.B. Leng, L. Yin, Z.T. Zhao, Q.G. Zhou

**SLAC:**  K. Bane, R. Hettel

**Soleil:**  J.-C. Denard, A. Nadji

**SPring-8:**  T. Fujita, H. Ohkuma, Y. Shimosaki, K. Soutome